\documentclass[%
 aip,
 amsmath,amssymb,
 reprint,
 ]{revtex4-1}

\usepackage{siunitx}
\usepackage{graphicx,color}
\usepackage{dcolumn}
\usepackage{bm}
\usepackage[normalem]{ulem}
\draft 

\begin{document}

\title{Tunable cavity-enhanced terahertz frequency-domain optical Hall effect}

\author{Sean Knight}
\email{sean.knight@engr.unl.edu}
\homepage{ellipsometry.unl.edu}
\affiliation{Department of Electrical and Computer Engineering, University of Nebraska-Lincoln, Lincoln,
NE, 68588-0511, USA}
\author{Stefan~Sch\"{o}che}
\affiliation{J.~A. Woollam Co., Inc., Lincoln, NE 68508-2243, USA}
\author{Philipp K\"{u}hne}
\affiliation{Terahertz Materials Analysis Center, THeMAC and Center for III-Nitride Technology, C3NiT–Janz\'{e}n, Department of Physics, Chemistry and Biology, Link\"{o}ping University, SE 58183 Link\"{o}ping, Sweden}
\author{Tino Hofmann}
\affiliation{Department of Physics and Optical Science, University of North Carolina at Charlotte, Charlotte, NC, 28223, USA}
\author{Vanya Darakchieva}
\affiliation{Terahertz Materials Analysis Center, THeMAC and Center for III-Nitride Technology, C3NiT–Janz\'{e}n, Department of Physics, Chemistry and Biology, Link\"{o}ping University, SE 58183 Link\"{o}ping, Sweden}
\author{Mathias Schubert}
\email{schubert@engr.unl.edu}
\affiliation{Department of Electrical and Computer Engineering, University of Nebraska-Lincoln, Lincoln,
NE, 68588-0511, USA}
\affiliation{Terahertz Materials Analysis Center, THeMAC and Center for III-Nitride Technology, C3NiT–Janz\'{e}n, Department of Physics, Chemistry and Biology, Link\"{o}ping University, SE 58183 Link\"{o}ping, Sweden}
\affiliation{Leibniz Institute for Polymer Research, 01069 Dresden, Germany}


\date{\today}

\begin{abstract}
Presented here is the development and demonstration of a tunable cavity-enhanced terahertz frequency-domain optical Hall effect technique. The cavity consists of at least one fixed and one tunable Fabry-P\'{e}rot resonator. The approach is suitable for enhancement of the optical signatures produced by the optical Hall effect in semi-transparent conductive layer structures with plane parallel interfaces. The physical principle is the constructive interference of electric field components that undergo multiple optical Hall effect induced polarization rotations upon multiple light passages through the conductive layer stack. Tuning one of the cavity parameters, such as the external cavity thickness, permits shifting of the frequencies of the constructive interference, and enhancement of the optical signatures produced by the optical Hall effect can be obtained over large spectral regions. A cavity-tuning optical stage and gas flow cell are used as examples of instruments that exploit tuning an external cavity to enhance polarization changes in a reflected terahertz beam. Permanent magnets are used to provide the necessary external magnetic field. Conveniently, the highly reflective surface of a permanent magnet can be used to create the tunable external cavity. The signal enhancement allows the extraction of the free charge carrier properties of thin films, and can eliminate the need for expensive super-conducting magnets. Furthermore, the thickness of the external cavity establishes an additional independent measurement condition, similar to, for example, the magnetic field strength, terahertz frequency, and angle of incidence. A high electron mobility transistor structure and epitaxial graphene are studied as examples. We discuss the theoretical background, instrument design, data acquisition, and data analysis procedures.
\end{abstract}

\maketitle 

\section{Introduction}\label{Sec:Introduction}

The optical Hall effect (OHE) is a phenomenon in which the optical response of a conductive material is altered by the presence of an externally applied magnetic field.\cite{SchubertJOSAA20_2003} This effect can be measured with generalized ellipsometry at oblique angles of incidence and at terahertz (THz) frequencies. Previously, the THz-OHE has been proven as a viable non-contact method to obtain the free charge carrier properties of semiconductor heterostructures using high-field superconducting magnets.\cite{KuehneRSI85_2014,HofmannAPL101_2012,HofmannTSF519_2011,HofmannAPL98_2011,SchocheAPL98_2011,Hofmannpssa205_2008,HofmannRSI77_2006} This approach allows the extraction of a sample's carrier concentration, mobility, and effective mass parameters by using a THz-transparent substrate as a Fabry-P\'{e}rot cavity to resonantly enhance the THz-OHE signal. Recently, it has been shown these properties can be conveniently obtained with permanent magnets.\cite{KnightOL40_2015,armakavicius2017cavity,KuehneIEEETHz2017,armakavicius2016properties} Using low-field permanent magnets to provide the external field significantly decreases the magnitude of the THz-OHE signal. However, one can compensate for this by exploiting an externally-coupled Fabry-P\'{e}rot cavity to further enhance the signal. In Ref.~\onlinecite{KnightOL40_2015}, different external cavity thickness values are achieved by simply stacking multiple layers of adhesive spacers between the sample and magnet. This method is useful because it is straight-forward and low cost, but only large increments of cavity thickness can be produced. The cavity-tuning optical stage described in this work improves on this previous approach and is capable of finely tuning the cavity thickness thus providing a new measurement dimension. As an example of the cavity-tuning stage described here a sample-permanent magnet arrangement is placed inside a gas flow cell to improve sensitivity to small variations of free charge carrier parameters under varying gas flow conditions. This experiment highlights the advantage of the small footprint of this enhancement technique.

In this work, we discuss the concept of THz-OHE signal enhancement due to an externally-coupled cavity. An optical model is used to choose desirable measurement parameters, such as angle of incidence, frequency, and external cavity thickness. Details of the instrument design and data acquisition are explained. Experimental and model-calculated data are presented and compared with data for the case of no cavity-enhancement. It is demonstrated that the cavity-enhancement technique allows extraction of the free charge carrier properties of a two-dimensional electron gas (2DEG) at THz frequencies.

\section{Method}\label{Sec:Method}

\subsection{Optical Hall effect}\label{subsec:OHE}

We refer to the OHE as a physical phenomenon that describes the occurrence of magnetic-field-induced dielectric displacement at optical wavelengths, transverse and longitudinal to the incident electric field, and analogous to the static electrical Hall effect.\cite{SchubertJOSAA20_2003,SchubertJOSAOHE2016} We have previously described data acquisition and analysis approaches for the method of the OHE in the mid-infrared, far-infrared, and THz spectral regions.\cite{SchubertJOSAA20_2003,HofmannRSI77_2006,KuehneRSI85_2014,KuehneIEEETHz2017} 

\subsection{Optical Hall effect model for thin film layer stacks}\label{subsec:OHEmodel}

The OHE can be calculated by use of appropriate physical models. The models contain two portions, one describes a given material's dielectric function under an external magnetic field, the second portion describes the wave propagation within a given layer stack. We have provided a recent review on this topic in Ref.~\onlinecite{SchubertJOSAOHE2016}. Briefly, in the THz spectral range the dielectric function can be approximated by a static contribution due to phonon excitations and higher energy electronic band-to-band transitions. Contributions due to free charge carriers can be well described by the Drude quasi-free electron model.\cite{Drude04,ANDP:ANDP19003060312} In the presence of a static external magnetic field, an extension of the Drude model predicts magneto-optic anisotropy\cite{DrudeAP1892} and which is the cause of the classical OHE in conductive materials. (For quantum effects see, for example, Refs.~\onlinecite{KuehnePRL111_2013,bouhafs2016decoupling,bouhafs2017multi})

\subsection{Tunable cavity-enhanced optical Hall effect}\label{subsec:TunablecavityenhancedopticalHalleffect}

\begin{figure*}[pbt]
\includegraphics[keepaspectratio=true,width=\linewidth]{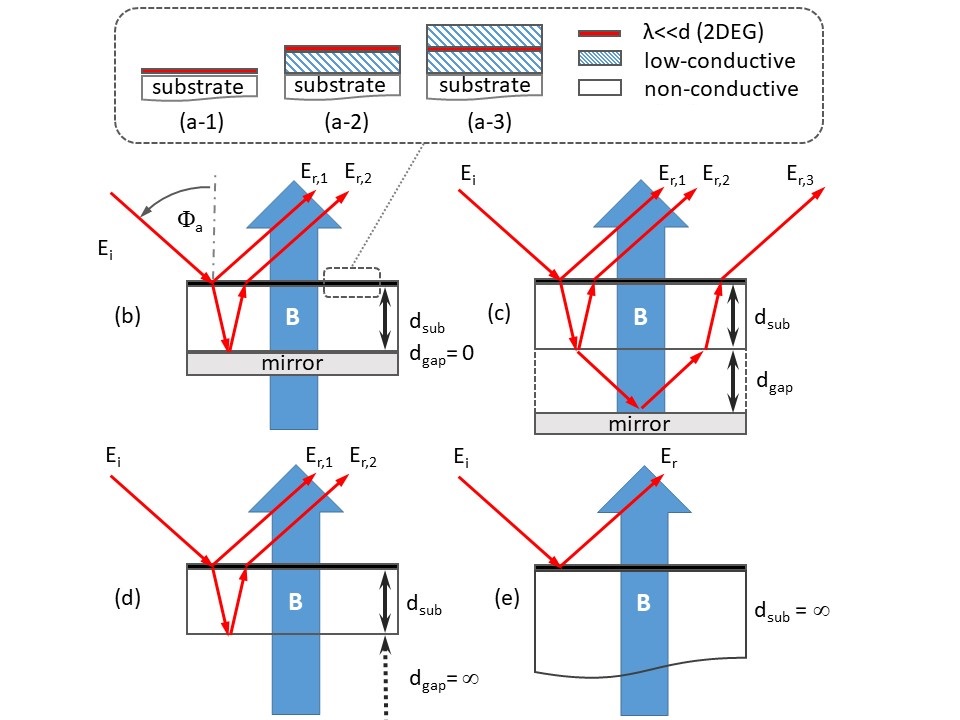}
\caption{Principle of the tunable cavity-enhanced frequency-domain THz-OHE method, here applied to characterize a two-dimensional electron gas (2DEG). The 2DEG of interest may be part of a multiple layer stack with differently, low-conducting constituents, for example, directly at the interface of a substrate (Fig.~\ref{fig:Sample-cavity_schematic}(a-1)), or ontop of the layer structure (Fig.~\ref{fig:Sample-cavity_schematic}(a-2)), or within (Fig.~\ref{fig:Sample-cavity_schematic}(a-3)). The differently, low-conducting constituents (layers) should themselves be sufficiently THz transparent. The principle configuration requires a THz-transparent substrate. If the substrate has a finite thickness ($d_{\mathrm{sub}}$), incident plane wave electric field ($E_i$) components are retro-reflected and pass the layer structure multiple times, where the first 2 orders are shown here for brevity only ($E_{r,1}, E_{r,2}$). A mirror placed at the opposite side of the substrate (Fig.~\ref{fig:Sample-cavity_schematic}b) can be used to control the frequencies of constructive interference maxima by $d_{\mathrm{gap}}$, where fractions of plane wave components reflected off the mirror exit the sample, and only the first order of those are shown ($E_{r,3}$). If the mirror surface is distanced to the backside of the substrate by $d_{\mathrm{gap}}$, the frequencies of constructive interference maxima can be tuned by $d_{\mathrm{gap}}$ (Fig.~\ref{fig:Sample-cavity_schematic}(c)). The angle of incidence is $\Phi_a$. Fig.~\ref{fig:Sample-cavity_schematic}(d) depicts the situation when the mirror is removed ($d_{\mathrm{gap}}\rightarrow \infty$). Fig.~\ref{fig:Sample-cavity_schematic}(e) depicts the case when the substrate is optically infinite ($d_{\mathrm{sub}}\rightarrow \infty$), and against which the enhancement of the tunable cavity-enhanced OHE is to be referenced. The magnetic field \textbf{B} direction is not relevant for the enhancement. Here, all examples are discussed with direction of \textbf{B} perpendicular to the cavity interfaces. Drawing not to scale.
\label{fig:Sample-cavity_schematic}}
\end{figure*}

The principle of the enhancement of the OHE in a layer stack is the constructive superposition of the magneto-optically polarization converted electromagnetic field components from multiple passages through the layer stack in the presence of an external magnetic field. Each time of passage, the electromagnetic field components undergo an additional polarization rotation caused by the magneto-optic anisotropy created by the response of the free charge carriers under the influence of the Lorentz force and within the conductive layer(s). At an interference maximum, the sensitivity to the portion of reflected light  that has undergone polarization discriminating reflection or transmission optical intensity measurement is greatly enhanced towards the causes of the magneto-optic anisotropy.\cite{KnightOL40_2015} The magnitude of the enhancement can be significant and which  depends on the free charge carrier properties in a given sample configuration. Examples are discussed in this work.

The principle of the tunable cavity-enhanced OHE method is demonstrated in Fig.~\ref{fig:Sample-cavity_schematic}. All configurations require the layer stack to be supported by a THz-transparent substrate. The substrate must have a flat and polished  backside whose surface is parallel to the front of the substrate carrying the layer stack. The thickness of the substrate $d_{\mathrm{sub}}$ should be such that spectrally neighboring Fabry-P\'{e}rot interference maxima and minima within the substrate can be sufficiently resolved with a given spectroscopic setup. Plane wave electric field ($E_i$) components incident under an angle $\Phi_a$ then pass the sample layer structure multiple times due to multiple internal reflections within the substrate cavity. The frequencies of such maxima are controlled by the angle of incidence, the substrate thickness and the substrate index of refraction. In principle, the substrate thickness is adjustable by depositing the sample layer stack onto different substrates. This, however, requires multiple fabrication steps. If a second cavity is created by the introduction of a mirror, placed at distance $d_{\mathrm{gap}}$, the portion of electromagnetic waves lost at the backside of the substrate is fed back into the substrate, and introduces additional fractions of plane wave components passing the layer stack. The two Fabry-P\'{e}rot cavities (substrate, gap) couple, and produce coupled Fabry-P\'{e}rot resonances. The frequencies of the coupled interference maxima can then be tuned by $d_{\mathrm{gap}}$, for any given but fixed $d_{\mathrm{sub}}$. Then, the magneto-optic signal enhancement occuring at interference maxima, limited to certain frequencies for a given $d_{\mathrm{sub}}$ without external cavity, can be tuned spectrally. Thereby, a new magneto-optic spectroscopy method is created where in addition to frequency, the external cavity is tuned by changing its thickness, $d_{\mathrm{gap}}$. 

Also shown in Fig.~\ref{fig:Sample-cavity_schematic} are the limiting cases, when the external cavity is zero ($d_{\mathrm{gap}}\rightarrow 0$, Fig.~\ref{fig:Sample-cavity_schematic}(b)), infinite ($d_{\mathrm{gap}}\rightarrow \infty$, Fig.~\ref{fig:Sample-cavity_schematic}(d)), and when both cavity and substrate are infinite ($d_{\mathrm{sub}}\rightarrow \infty$, Fig.~\ref{fig:Sample-cavity_schematic}(e)). The case $d_{\mathrm{gap}}\rightarrow 0$ requires deposition of a metal layer onto the backside of the substrate. The case $d_{\mathrm{gap}}\rightarrow \infty$ occurs when the substrate is THz-transparent and has parallel interfaces. The case $d_{\mathrm{sub}}\rightarrow \infty$ occurs when the layer stack is deposited onto a non-transparent substrate or when the backside of a transparent substrate is not parallel, for example, if the substrate consists of a wedge or a prism. 

\subsection{Mueller matrix spectroscopic ellipsometry}\label{subsec:MMSE}

The generalized ellipsometry concept\cite{Azzam72}, its spectroscopic extension,\cite{SchubertJOSAA13_1996} and the Mueller matrix formalism\cite{JarrendahlMuellerJANews2011} are employed in this work. The Mueller matrix connects Stokes vector\cite{JonesJOSA1947} components of electromagnetic waves before and after interaction with the sample upon reflection or transmission. For the use of the Mueller matrix concept in spectroscopic generalized ellipsometry we refer the reader to recent reviews (see, e.g., Refs.~\onlinecite{SchubertIRSEBook_2004,Fujiwara_2007}). For use of the Mueller matrix formalism in the OHE\cite{footnoteSeanRSI2017A}, and in particular, for data format definition we refer to Ref.~\onlinecite{KuehneRSI85_2014}.

\subsection{Data analysis}\label{subsec:dataanalysis}

Non-linear parameter regression analysis methods are used for data analysis. The experimental data are compared with calculated OHE data. The calculated data are obtained with appropriate physical models and model parameters. Parameters are varied until a best-match is obtained minimizing an appropriately weighted error sum. The error sum takes into account the systematic uncertainties determined during the measurement for each experimental data value. Best-match model parameter uncertainties are obtained from the covariance matrix using the 90$\%$ confidence interval.\cite{KuehneRSI85_2014}

\section{Instrument}\label{Sec:Instrument}

The tunable cavity and sample must be placed within a THz spectroscopic ellipsometer system, and subjected to an external magnetic field. Ellipsometer, sample stage, and magnetic field designs are discussed in this section. 

\subsection{Terahertz frequency-domain ellipsometer}\label{subs:ecTHzOHE}

Two THz frequency-domain ellipsometer instruments are used in this work. Both instruments operate in the rotating-analyzer configuration which enables acquisition of the upper left 3$\times$3 block of the 4$\times$4 Mueller matrix. The frequency-domain source is a backward wave oscillator (BWO) with GaAs Schottky diode frequency multipliers. Technical details are described in Ref.~\onlinecite{KuehneRSI85_2014} and Ref.~\onlinecite{KuehneIEEETHz2017}. 

\subsection{Tunable cavity stage}\label{subsec:tunablecavitystage}

\begin{figure}
\includegraphics[keepaspectratio=true,width=\linewidth, clip, trim=0cm 0cm 0cm 0cm ]{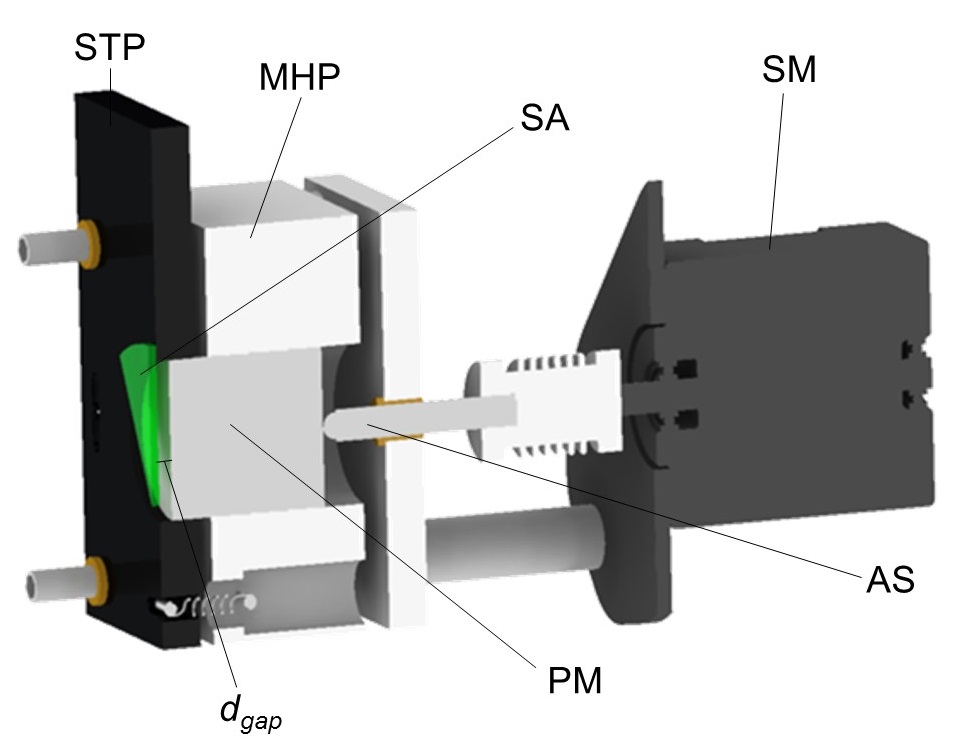}
\caption{Technical schematic of the sample holder with cavity-tuning adjustment controls. The  instrument is comprised of the spring-loaded sample tip-tilt plate (STP), the sample (SA), the stepper motor (SM), the mirror housing plate (MHP), the adjustment screw (AS), and the permanent magnet (PM). The external cavity distance (sample-magnet air gap) is labeled $d_{\text{gap}}$. The plate containing the AS is removable which allows the user to flip the permanent magnet to the opposite pole-face and redo experiments with opposite field direction without disturbing the sample alignment.} \label{fig:optical_stage}
\end{figure}

\paragraph{Use of permanent magnet:} A principle design of the sample holder for the tunable cavity-enhanced THz-OHE is shown in Fig.~\ref{fig:optical_stage}. In this design, the permanent magnet serves both as mirror as well as for providing the external magnetic field. The mirror properties of the magnet surface must be characterized by THz spectroscopic ellipsometry measurements at multiple angles of incidence prior to its use in the sample stage. The permanent magnet (PM) sits flush inside a hole in the mirror housing plate (MHP). The micrometer adjustment screw (AS) rests inside a brass bushing in the back plate. The rounded tip of the AS is made of ferromagnetic material which attracts the magnet providing synchronous PM-AS movement. The stepper motor (SM) is attached to the back of the setup and is connected to the AS by a flexible bellows shaft coupler. The flexible coupler allows a $d_{\text{gap}}$ range of approximately 0~$\mu$m to 600~$\mu$m. The stepper motor is operated by a commercially available motor controller (Thor Labs Inc.), which uses LabVIEW programming. The minimum $d_{\text{gap}}$ increment is 1.6~$\mu$m, corresponding to one step of the motor. The back plate in Fig.~\ref{fig:optical_stage} is removable and allows the user to flip the magnet to the opposite pole-face and redo experiments without disturbing the sample alignment.

\paragraph{Use of fixed cavity spacer adjustments:} For simplifying the tunable cavity-enhanced sample stage, non-magnetic adhesive spacers can be placed between the sample and the mirror surface dispensing with the need for the stepper motor in Fig.~\ref{fig:optical_stage}. This option is suitable for \textit{in-situ} measurements when limited space is available. However, no tuning of the cavity after sample mounting can be performed.  
\paragraph{Use of external electromagnet:} For use of the sample holder with an external electromagnet, the permanent magnet can be replaced by a non-magnetic insert with a THz mirror at the front towards $d_{\mathrm{gap}}$ and the sample backside. The normal reflectance properties of the mirror can be evaluated by performing THz spectroscopic ellipsometry measurements at multiple angles of incidence prior to its use. The external magentic field can be provided by electromagnets, for example, by placing the stage within a Helmholtz coil arrangement.
   
\section{Data Acquisition and Analysis}\label{Sec:Dataacquisitionandanalysis}

\subsection{Data acquisition}\label{subsec:dataaquisition}

\paragraph{Magnetic field calibration:} The permanent magnet mounted in the sample holder is a high-grade neodymium (N42) magnet. With the use of a permanent magnet the change in magnetic field strength at the sample surface upon variation of $d_{\text{gap}}$ can be substantial. Hence, it is necessary to implement the magnetic field as a function of distance in the optical model. Using a commercially available Hall probe (Lakeshore), the magnetic field is measured at multiple $d_{\text{gap}}$ values. For our instrument, within approximately 1~mm of the magnet surface the field is approximately linear and can be approximated using by:
\begin{equation}
\pm B = \pm[0.55 - (5.1\times10^{-5})\times (d_{\text{gap}}+d_{\text{sub}})]~\text{[T]},
\end{equation}
where the plus and minus sign refers to the two respective pole orientations of the magnet. The parameters $d_{\text{gap}}$ and $d_{\text{sub}}$ are in units of micrometers.

\paragraph{Mirror calibration:} Separate ellipsometry experiments are performed in the mid-infrared spectral range to determine the optical properties of the metallic permanent magnet surface as mirror. Data analysis is performed using the classical Drude model parameters for static resistivity of $\rho = (9.53\pm 0.04) \times 10^{-5}~\Omega$cm and the average-collision time $\tau = (1.43\pm0.08) \times 10^{-16}~$s. These parameters are used here to model-calculate the optical reflectance of the magnet surface for the model analysis in the THz spectral range for the cavity-enhanced measurements. The magnet surface behaves as an ideal metallic ``Drude'' mirror characterized by metal electron carrier scattering time and resistivity, and no magneto-optic polarization coupling occurs because the metal electron effective mass is too large and the mean scattering time is too short in order for the free charge carriers to respond to the external magnetic field producing measurable magnetooptic birefringence.  

\paragraph{Mirror-to-ellipsometer alignment:} The mirror surface is aligned first and then the sample is mounted and aligned. The mirror surface is aligned to the ellipsometer's coordinate system by use of a laser diode mounted such that the laser diode beam is parallel to the plane of incidence, perpendicular to the sample surface, and coincides with the center of the THz beam at the sample surface. A gap value $d_{\text{gap}}$ is selected in the middle of the range of values anticipated for experiments. To align the mirror, the alignment laser diode beam is reflected off the mirror surface and the mirror is adjusted until the beam reflects back into the laser aperture. The adjustment is performed by moving the entire stage relative to the ellipsometer system.

\paragraph{Sample-to-mirror alignment:} Once the mirror is aligned, the sample is mounted to the sample tip-tilt plate STP (Fig.~\ref{fig:optical_stage}). STP serves as an adjustable frame to mount the sample. The sample can be mounted via adhesive, for example, or mechanical clamps. STP contains three micrometer screws secured against the MHP by springs, creating a tip-tilt ability. This is necessary to ensure that the sample surface is also aligned to the ellipsometer's coordinate system. The sample surface is aligned using the same alignment laser as for the mirror.

\begin{figure}
\includegraphics[keepaspectratio=true,width=\linewidth, clip, trim=0cm 0cm 0cm 0cm ]{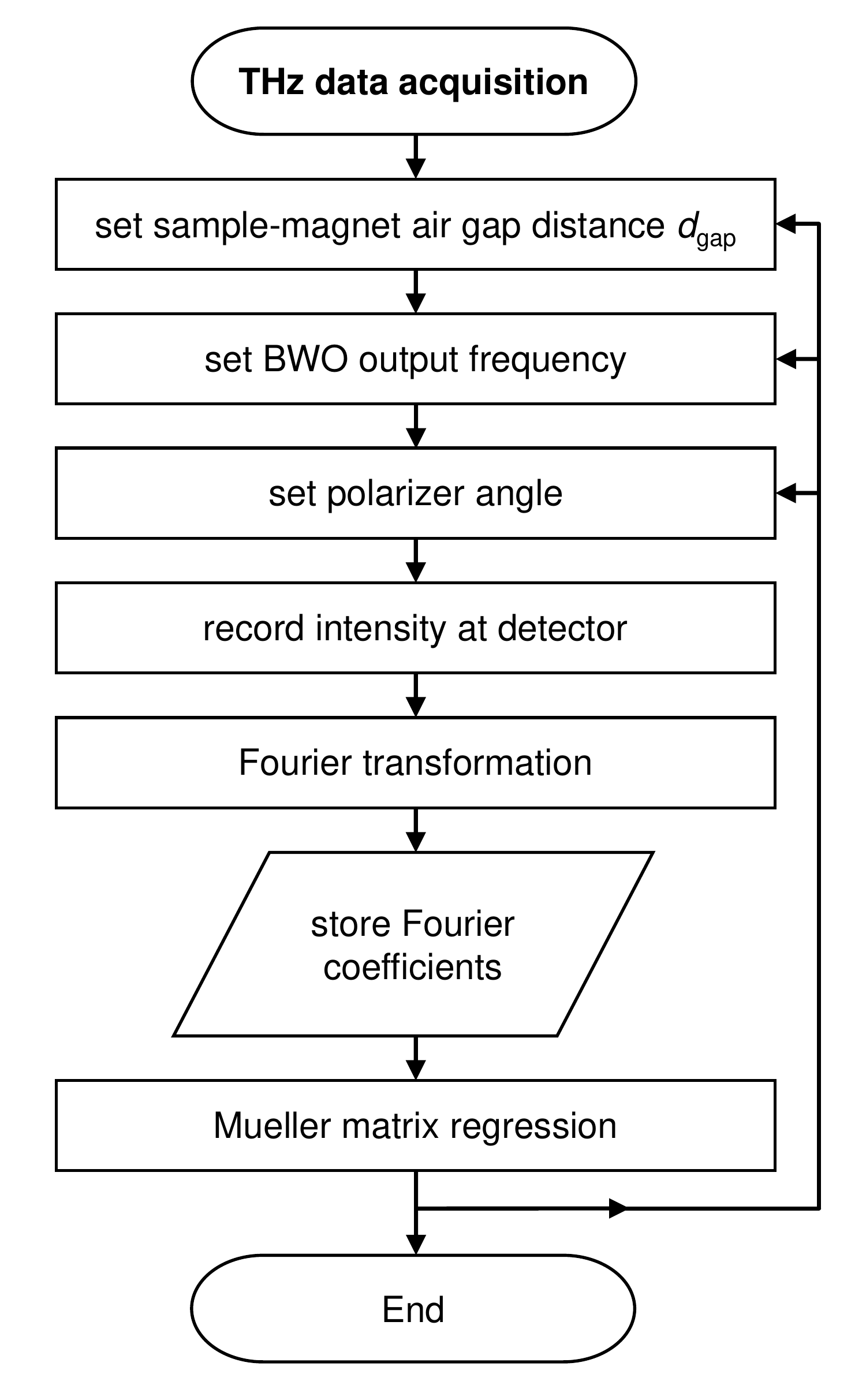}
\caption{Flow chart describing the data acquisition process of the cavity-tuning optical stage and ellipsometer.
\label{fig:flow_chart}}
\end{figure}

\paragraph{Ellipsometry data acquisition:} After mounting the sample stage into the ellipsometer system, data are acquired in a selected spectral range, for selected angles of incidence $\Phi_a$, and gap distance $d_{\mathrm{gap}}$. Figure~\ref{fig:flow_chart} depicts a flow chart describing the data acquisition process. First, the sample-mirror air gap distance $d_{\text{gap}}$ is set. Next, the frequency-domain source frequency is set. Then the polarizer angle is set and the intensity at the detector is recorded. The process is repeated for all polarizer settings as described in Ref.~\onlinecite{KuehneRSI85_2014}, and a Fourier transform of the signal is performed to determine the Fourier coefficients, which are stored and then subject to a regression analysis. As a result, the elements of the upper 3$\times$3 block of the Mueller matrix are obtained. The procedure is repeated for different settings of gap distance, frequency, or angle of incidence, for example. The acquisition process can be repeated with the magnetic field direction reversed, for example, by reverting the magnet direction, or by reverting the currents in external electromagnetic coils. The experiment can also be repeated with a mirror without a magnet for acquisition of field-free ellipsometry data.

\subsection{Data analysis}\label{subsec:dataanalysis}

Data measured by tunable cavity-enhanced THz-OHE are analyzed using model calculations and numerical regression procedures. Multiple data acquisition modes are available, and which will be discussed by examples further below.

\paragraph{Cavity-enhanced data at tunable gap thickness:} Data obtained as a function of gap thickness are compared with calculated data.

\paragraph{Cavity-enhanced data at tunable frequency:} Data obtained at a fixed gap and/or substrate thickness but as a function of frequency are compared with calculated data.

\paragraph{Cavity-enhanced data at tunable frequency and tunable gap thickness:} Data over a 2-dimensional parameter set can be obtained tuning both gap thickness and frequency and are compared with calculated data.

\paragraph{Cavity-enhanced data at magnetic field reversal:} Field-reversal OHE data obtained at opposing magnetic field directions, $\Delta M_{ij} = M_{ij}(B) - M_{ij}(-B)$, are taken and the difference data is compared with calculated data.

\section{Results and Discussion}\label{Sec:Resultsanddiscussion}

Here we discuss two sample systems as examples for the application of the tunable cavity-enhanced OHE. Both samples contain 2DEGs. The characterization of their free charge carrier properties is demonstrated.  One sample is comprised of a transistor device structure for a high electron mobility transistor (HEMT) based on group-III nitride semiconductor layer structures. The second sample is an epitaxial graphene sample grown on a silicon carbide substrate.

\subsection{Two-dimensional electron gas characterization in a HEMT device structure}\label{subsec:HEMT}

\subsubsection{Sample structure}\label{sec:samplestructure}

\paragraph{Growth:} The sample investigated is an AlInN/AlN/GaN HEMT structure grown using an AIXTRON 200/4 RF-S metal-organic vapor phase epitaxy system. The HEMT structure consists of a bottom 2~$\mu$m thick undoped GaN buffer layer, a 1~nm thick AlN spacer layer, followed by a 12.3~nm thick Al$_{\text{0.82}}$In$_{\text{0.18}}$N top layer.\cite{DarakchievaJAP08,schoeche2014_phd-dissertation} The substrate is single-side polished \textit{c}-plane sapphire with a nominal thickness of 350~$\mu$m. 

\paragraph{Optical sample structure:} All sample constituents are optically uniaxial and the layer interfaces are plane parallel. In a separate experiment, the HEMT structure was investigated using a commercial (J.A. Woollam Co. Inc.) mid-infrared (MIR) ellipsometer from 300-1200~cm$^{-1}$ at $\Phi_a$~=~60$^{\circ}$ and 70$^{\circ}$ at room temperature in order to determine phonon mode parameters of the AlInN top layer.  No distinct phonon features are seen in the THz measurements. However, the MIR analysis is used to help determine the dielectric function of the HEMT structure constituents in the THz spectral range. Phonon parameters for the substrate, GaN buffer layer, and AlN spacer layer are taken from Ref. \onlinecite{SchubertPRB61_2000}, \onlinecite{SchocheAPL98_2011}, and \onlinecite{SchocheAPL103_2013}, respectively. The thickness of the AlInN and AlN layers are found by growth rate calculations and not varied in the analysis. The best-match model layer thickness for the GaN layer is $(2.11\pm0.01)~\mu$m. For the AlInN top layer, the best-match model frequency and broadening parameters for the one-mode type E1- and A1-symmetry are 
$\omega_{\text{TO},\perp}=(625.4\pm0.8)~\text{cm}^{-1}$, $\omega_{\text{LO},\perp}=(877.8)~\text{cm}^{-1}$, $\gamma_{\perp}=(40.8\pm1.5)~\text{cm}^{-1}$,
$\omega_{\text{TO},\parallel}=(610)~\text{cm}^{-1}$, $\omega_{\text{LO},\parallel}=(847.8\pm0.4)~\text{cm}^{-1}$, $\gamma_{\parallel}=(11.3\pm0.4)~\text{cm}^{-1}$
which are in good agreement with previous works\cite{KasicPSSC_2003}. Note, certain phonon parameters are functionalized according to Ref. \onlinecite{KasicPSSC_2003}, and were not varied in the analysis. In order to obtain an excellent match between experimental and model-calculated THz-OHE data, a low-mobility electron channel was included in the AlInN top layer. This same low-mobility channel was also included in our previous model analysis for the same HEMT structure.\cite{schoeche2014_phd-dissertation,KnightOL40_2015} A mobility value of $\mu =50~\text{cm}^{2}/\text{Vs}$ for a similar HEMT structure is adopted for this sample, and an effective mass parameter of 0.3~$m_{0}$ is taken from density function calculations in Ref.~\onlinecite{tsai2019structural}. The volume density value for the low-mobility channel was previously determined to be $N = 1.02 \times 10^{20}~\text{cm}^{-3}$. This value is not varied in our analysis.

\paragraph{Previous OHE characterization:} In Ref.~\onlinecite{schoeche2014_phd-dissertation} we reported field-reversal high-field OHE measurements on the same HEMT sample without external cavity. The high field measurements were performed in a cryogenic superconducting magnet setup. In Ref.~\onlinecite{KnightOL40_2015} we reported field-reversal cavity-enhanced OHE measurements using a permanent magnet and various adhesive spacers (discrete settings for $d_{\mathrm{gap}}$) on the same sample. The results reported in this work are in excellent agreement with those reported previously. All THz-OHE results further compare well with Hall effect and C-V measurements done on similar samples.\cite{GonschorekJAP103_2008,GonschorekAPL89_2006}

\subsubsection{Single-frequency tunable-cavity measurements}

\begin{figure}
\includegraphics[keepaspectratio=true,width=\linewidth, clip, trim=0cm 0cm 0cm 0cm ]{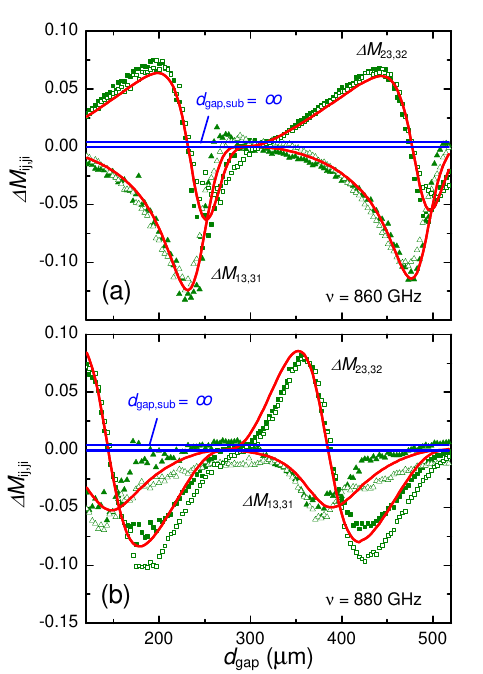}
\caption{Experimental (green symbols) and best-match model calculated data (red solid lines) field-reversal cavity-enhanced OHE data as a function of the external cavity thickness $d_{\mathrm{gap}}$, at two frequencies for a HEMT layer structure on sapphire. Closed and open triangles represent $\Delta M_{\text{13}}$ and $\Delta M_{\text{31}}$ respectively. Closed and open squares represent $\Delta M_{\text{23}}$ and $\Delta M_{\text{32}}$ respectively. The blue solid lines are model-calculated data for the case of no cavity-enhancement ($d_{\text{gap}}=\infty$, $d_{\text{sub}}=\infty$). All data is obtained at angle of incidence $\Phi_a = 45^{\circ}$ and at room temperature.
\label{fig:line_scans}}
\end{figure}

Figure~\ref{fig:line_scans} shows experimental and best-match model data for field-reversal cavity-enhanced OHE data as a function of $d_{\text{gap}}$ for the HEMT sample. The experiment was performed at two different, fixed frequencies of $\nu$~=~860~GHz and 880~GHz for a $d_{\text{gap}}$ range of 120~$\mu$m to 520~$\mu$m in increments of 3~$\mu$m. The experimental data for both frequencies are analyzed simultaneously. The layer stack optical model for the best-match model calculation is AlInN/AlN/GaN/sapphire substrate/external cavity/mirror (magnet surface). The external magnetic field is oriented normal to the sample surface. All off-block diagonal Mueller matrix elements are zero for the HEMT sample structure without external magnetic field. To begin with, the solid blue lines in Figs.~\ref{fig:line_scans}(a,b) are model calculated data for the same HEMT structure in the absence of the cavity enhancement (Fig.~\ref{fig:Sample-cavity_schematic}(e), $d_{\text{sub}} = \infty$, $d_{\text{gap}} = \infty$), where we assumed that the field at the layer stack is $B=\pm 0.55$~T. Specifically, $\Delta M_{\text{13}}=\Delta M_{\text{31}}=-0.0004$ and $\Delta M_{\text{23}}=\Delta M_{\text{32}}=0.004$. Data are below our current instrumental uncertainty limit for the individual Mueller matrix elements of $\delta M_{ij} \approx \pm 0.01$. Hence, the 2DEG within the HEMT layer structure would not be detectable. 
In the cavity-enhanced mode, however, large off-diagonal Mueller matrix elements appear, far above the current instrumental uncertainty level, upon variation of the gap thickness $d_{\text{gap}}$. Features in Figs.~\ref{fig:line_scans}(a,b) are due to Fabry-P\'{e}rot interference enhanced cross-polarized field components after reflection at the layer stack. Minima and maxima occur as a function of gap thickness. The cross polarization is produced only by the free charge carrier gas within the HEMT structure under the influence of the Lorentz force. The potential to use the variation of gap thickness as a new parameter variation measurement configuration is obvious. In particular, for this sample, and for the scans shown in Figs.~\ref{fig:line_scans}(a,b), compared with the blue lines (no cavity enhancement) cavity thickness parameters can be adjusted where the OHE signal enhancement reaches -0.124 at $\nu$=860 GHz in Fig.~\ref{fig:line_scans}(a) for $\Delta M_{\text{13,31}}$ and 0.088 for $\nu$=880 GHz in Fig.~\ref{fig:line_scans}(b) for $\Delta M_{\text{23,32}}$. The best-match model 2DEG sheet density, mobility, and effective mass parameters obtained from the OHE data in Figs.~\ref{fig:line_scans}(a,b) are $N_\text{s} = (1.23\pm0.13) \times 10^{13}~\text{cm}^{-2}$, $\mu = (1245\pm64)~\text{cm}^{2}/\text{Vs}$, $m^{\ast} = (0.272\pm0.013) m_{0}$, respectively. The results are in excellent agreement with electrical measurements and with our previous THz-OHE experiments.\cite{KnightOL40_2015}

\subsubsection{Tunable-frequency tunable-cavity measurements}

\begin{figure}
\includegraphics[keepaspectratio=true,width=\linewidth, clip, trim=0cm 0cm 0cm 0cm ]{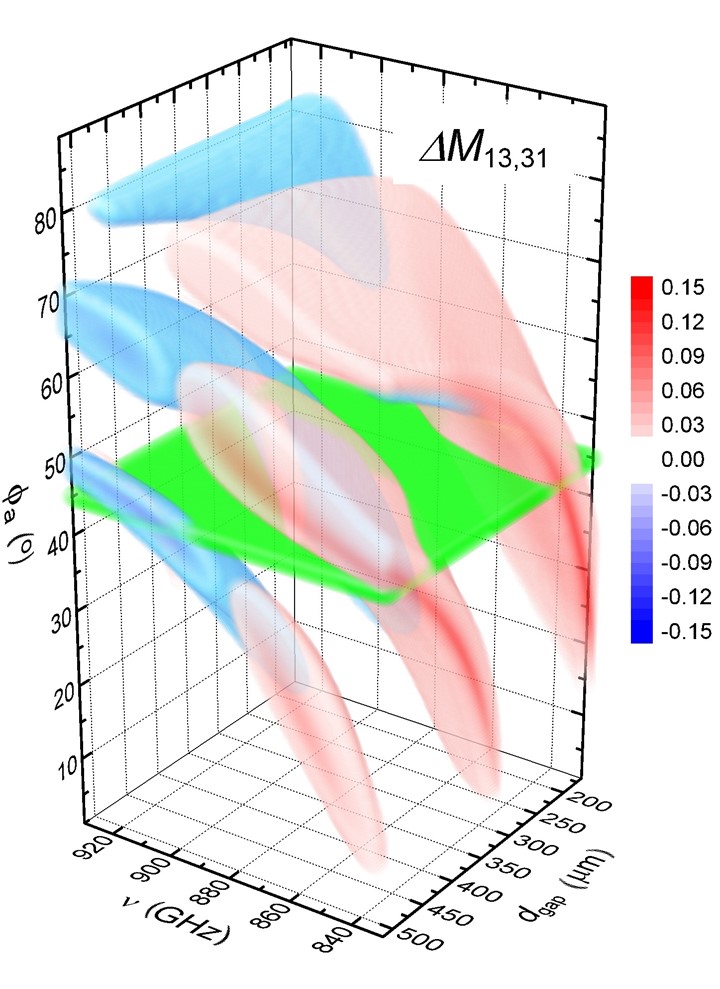}
\caption{False-color three-dimensional surface rendering of model-calculated cavity-enhanced field-reversal THz-OHE data for an AlInN/AlN/GaN HEMT structure grown on a sapphire substrate as functions of frequency $\nu$, external cavity distance $d_{\text{gap}}$, and angle of incidence $\Phi_a$. Data for $\Delta M_{13,31} = M_{13,31}(+B) - M_{13,31}(-B)$ are shown as example. Values within the range of -0.02 to 0.02 are omitted for clarity. The green horizontal plane at $\Phi_a = 45^{\circ}$ indicates the instrumental  settings for the angle of incidence in this work. Model parameters given in text. Note that the model calculated plot for $\Delta M_{23,32}$ is similar in appearance to $\Delta M_{13,31}$ and is excluded here for brevity.
\label{fig:M13-31_3D}}
\end{figure}

\begin{figure*}
\includegraphics[keepaspectratio=true,width=\linewidth, clip, trim=0cm 0.2cm 0cm 0.2cm ]{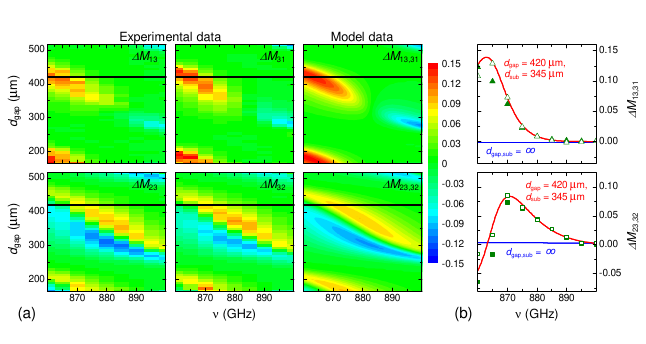}
\caption{(a) False-color two-dimensional surface rendering of experimental (left two columns) and model-calculated (right column) cavity-enhanced field-reversal THz-OHE data for an AlInN/AlN/GaN HEMT structure grown on a sapphire substrate as functions of frequency $\nu$ and external cavity distance $d_{\text{gap}}$. The angle of incidence is $\Phi_a = 45^{\circ}$. Data for $\Delta M_{13,31} = M_{13,31}(+B) - M_{13,31}(-B)$ are shown as example. Parameter details given in text. (b) Same as in (a) for fixed cavity thickness $d_{\text{gap}}$~=~420~$\mu$m. The solid blue line indicates model calculated data when $d_{\text{sub}}=\infty$ and $d_{\text{gap}}=\infty$. Solid green symbols indicate $\Delta M_{\text{13}}$ and $\Delta M_{\text{23}}$, and open symbols indicate $\Delta M_{\text{31}}$ and $\Delta M_{\text{32}}$. The blue solid lines are model-calculated data for the case of no cavity-enhancement ($d_{\text{gap}}=\infty$, $d_{\text{sub}}=\infty$). All data taken at room temperature. \label{fig:OHE_data}}
\end{figure*}

False-color rendering of model calculated data of cavity-enhanced field-reversal THz-OHE data versus frequency, gap thickness, and angle of incidence are shown in Fig.~\ref{fig:M13-31_3D}. The color type indicates positive or negative values, the color intensity indicates the magnitude of the OHE data. The three-dimensional rendering is insightful as it indicates distinct regions within which the data rapidly switches signs, and regions within which data takes very large values. All three parameters, frequency, gap thickness, and angle of incidence influence the OHE data, and proper selection may result in strong OHE data, while poor choices may result in disappearance of the OHE data. The horizontal plane indicated at 45$^{\circ}$ identifies the angle of incidence at which experiments are performed in this work. False-color rendering of experimental and model-calculated data at angle of incidence of $45^{\circ}$ are shown in Fig.~\ref{fig:OHE_data}(a) as a function of frequency and gap thickness. An excellent agreement between both experiment and model calculation is obtained. Fig.~\ref{fig:OHE_data}(b) shows data at a fixed cavity thickness. Data are similar to those in Figs.~\ref{fig:line_scans}(a,b), except now the frequency is tuned. Figs.~\ref{fig:OHE_data}(a,b) identify frequency and gap regions where the OHE data is very small. The blue solid lines are identical to those in Figs.~\ref{fig:line_scans}(a,b) for the case of no cavity-enhancement. All experimental data in Fig.~\ref{fig:OHE_data} is analyzed simultaneously and the resulting best-model sheet density, mobility, and effective mass parameters for the 2DEG are $N_\text{s} = (1.22\pm0.12) \times 10^{13}~\text{cm}^{-2}$, $\mu = (1262\pm59)~\text{cm}^{2}/\text{Vs}$, $m^{\ast} = (0.268\pm0.012) m_{0}$. The results are identical within the error bars to those obtained from the single frequency gap thickness scans as well as to our previous OHE investigation reports.\cite{KuehneRSI85_2014,KnightOL40_2015} 

The central results obtained from this section are (i) the demonstration of the strong enhancement obtained by use of multiple interference through substrate and external cavities, and (ii) the obvious appearance from Fig.~\ref{fig:M13-31_3D} and the support from our experiments that OHE measurements can be performed both as a function of frequency and gap thickness.  

\subsection{Environmental gas doping characterization in epitaxial graphene}\label{subsec:gasdoping}

In this section we demonstrate the use of the cavity-enhanced OHE method for detecting changes in the properties of a 2DEG upon exposure to various external gas compositions. The purpose of this section is to demonstrate the use of this method when transient physical changes to a sample limit the time durations during which spectroscopic scanning measurements can be performed. 

\subsubsection{Sample structure}

\paragraph{Growth:} The sample studied here is graphene epitaxially grown on Si-face (0001) 4H-SiC by high-T sublimation in Ar atmosphere.\cite{knight2017situ} Reflectivity and low-energy electron microscopy mapping, and scan lines verify the primarily one monolayer coverage across the 10$\times$10 mm sample surface.

\paragraph{Optical sample structure:} The sample is optically modeled by considering the graphene monolayer as a 1~nm highly conductive thin film on top of the SiC substrate as described in Ref.~\onlinecite{knight2017situ}. All sample constituents have plane parallel interfaces. No free charge carriers are detected in the SiC substrate. Due to the ultrathin layer thickness of the graphene, ellipsometry data cannot differentiate between the thickness and the dielectric function of the layer. Instead, a new parameter emerges, the sheet free charge carrier density,  This parameter takes a constant ratio with the assumed layer thickness, and hence can be determined independently and accurately (Further details are discussed in  Refs.~\onlinecite{knight2017situ, SchubertJOSAOHE2016}).

\paragraph{\textit{In-situ} gas cell design:} The schematics of the \textit{in-situ} gas flow cell used in this work is shown in Fig.~\ref{fig:gas-cell-experiments}(a). The THz ellipsometer instrument is schematically indicated by source, polarizer, analyzer, and detector at angle of incidence $\Phi_a$~=~45$^{\circ}$.\cite{KuehneIEEETHz2017} The cell is equipped with a humidity and temperature sensor, gas inlets, and gas outlets. The side walls of the flow cell are made from Delrin, and the cover and base portions are made from acrylic. THz-transparent windows are produced from homopolymer polypropylene. The thickness of the transparent sheets is 0.27~mm. Normal ambient gas is pushed through the cell using a vacuum pump (Linicon). Nitrogen and helium flow was provided by additional purge lines. The background pressure in the cell was 1~atm throughout the experiment. The flow rate was 0.5~liters/minute. 

\paragraph{Fixed cavity-enhancement settings:} The sample consists of a 2DEG (graphene) at the surface of a THz-transparent substrate ($d_{\mathrm{sub}}$~=~355~$\mu$m). The substrate is placed with its backside using adhesive spacers onto the permanent magnet (Fig.~\ref{fig:gas-cell-experiments}(a)). The sample is mounted with the neodymium (N42) magnet into the gas cell. The gap thickness $d_{\text{gap}}$ was thereby fixed at 100~$\mu$m. OHE data acquisition is identical to the procedure in Fig.~\ref{fig:flow_chart} with fixed cavity thickness.

\subsubsection{Cavity-enhanced optical Hall effect simulations}

\begin{figure}
\includegraphics[keepaspectratio=true,width=\linewidth, clip, trim=0cm 0cm 0cm 0cm ]{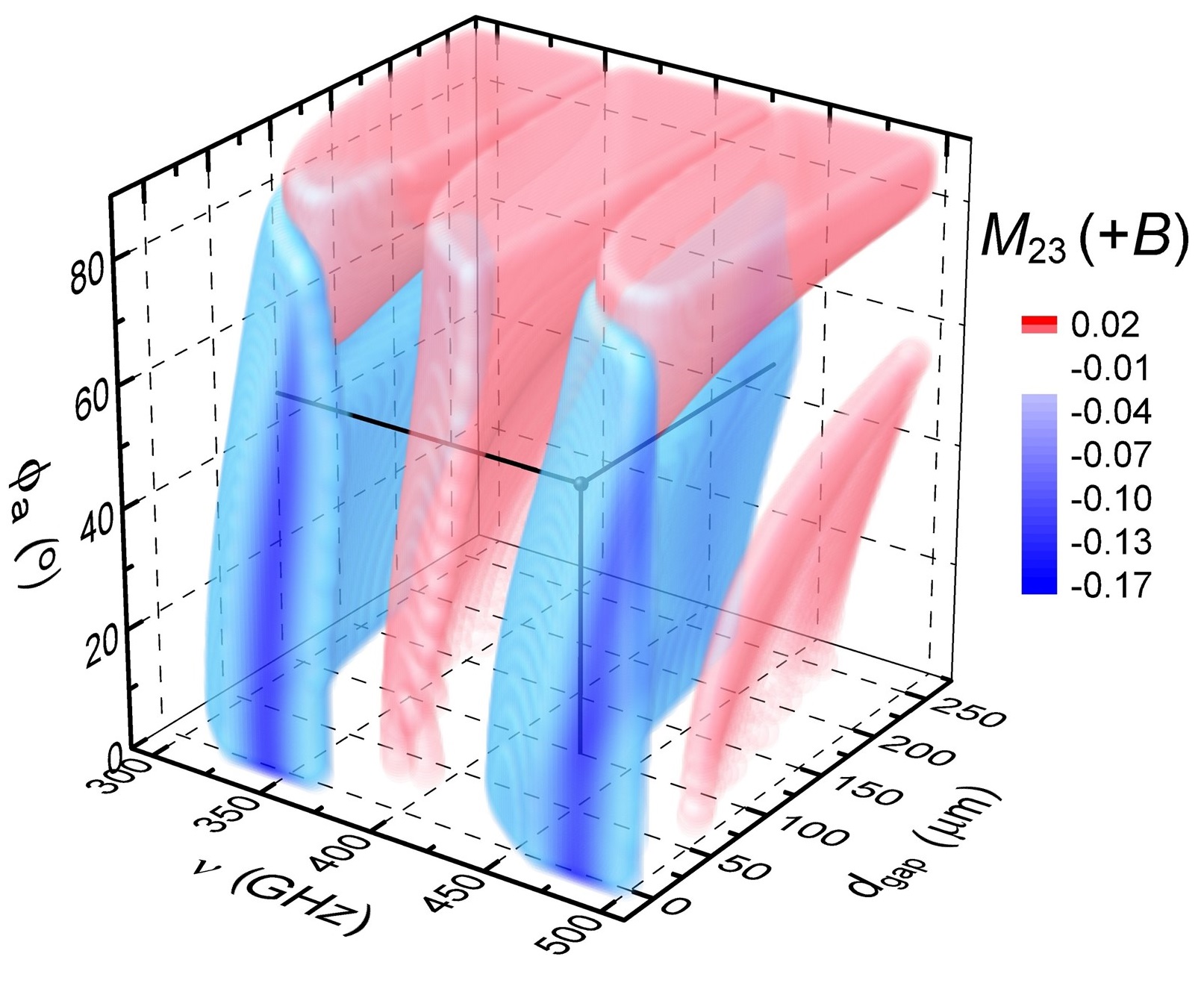}
\caption{False-color three dimensional surface rendering of model-calculated cavity-enhanced single-field THz-OHE data for an epitaxial graphene layer on SiC as a function of frequency $\nu$, external cavity distance $d_{\text{gap}}$, and angle of incidence $\Phi_a$. Data for $M_{23}(+B)$ are shown only (0.02 to 0 (red) and -0.04 to -0.17 (blue)). Data with values between -0.04 and 0 are omitted for clarity. The black sphere at the intersection of the three black lines illustrates the point where \textit{in-situ} gas cell data was taken for the gas flow experiments on the sample shown on this work. For model calculations the following parameters are used: $N_\text{s} = 8.69\times10^{11}~\text{cm}^{-2}$ and $\mu = 2550~\text{cm}^{2}/\text{Vs}$. The effective mass parameter of $m^{\ast}$~=~0.019~$m_0$ is calculated as a function of $N_\text{s}$ as in Ref.~\onlinecite{novoselov2005two}.}
\label{fig:graphene_m23_32_4d}
\end{figure}

Figure~\ref{fig:graphene_m23_32_4d} shows model-calculated THz-OHE data for $M_{23}$ as function of $\nu$, $d_{\text{gap}}$, and $\Phi_a$ for epitaxial graphene grown on SiC. Figure~\ref{fig:graphene_m23_32_4d} can be used as a guide to find optimal values for $\nu$, $d_{\text{gap}}$, and $\Phi_a$ to perform a THz-OHE measurement. The black sphere and three intersecting lines illustrates the point chosen to perform the \textit{in-situ} THz-OHE measurement. Only a single set of measurement parameters is chosen to minimize time between measurements in order to resolve sharp dynamic changes in the Mueller matrix data during the gas flow experiment. For practical reasons, $d_{\text{gap}} = 100~\mu$m and $\Phi_a = 45^{\circ}$ were chosen. Therefore, $\nu$ = 428~GHz was selected for the \textit{in-situ} measurement. Unlike Fig.~\ref{fig:M13-31_3D}, Fig.~\ref{fig:graphene_m23_32_4d} is not THz-OHE difference data since the gas flow experiment is only performed using the north pole-face of the permanent magnet.

\subsubsection{\textit{In-situ} tunable-frequency single-cavity measurements}

\begin{figure*}
\includegraphics[keepaspectratio=true,width=\linewidth, clip, trim=0cm 0cm 0cm 0cm ]{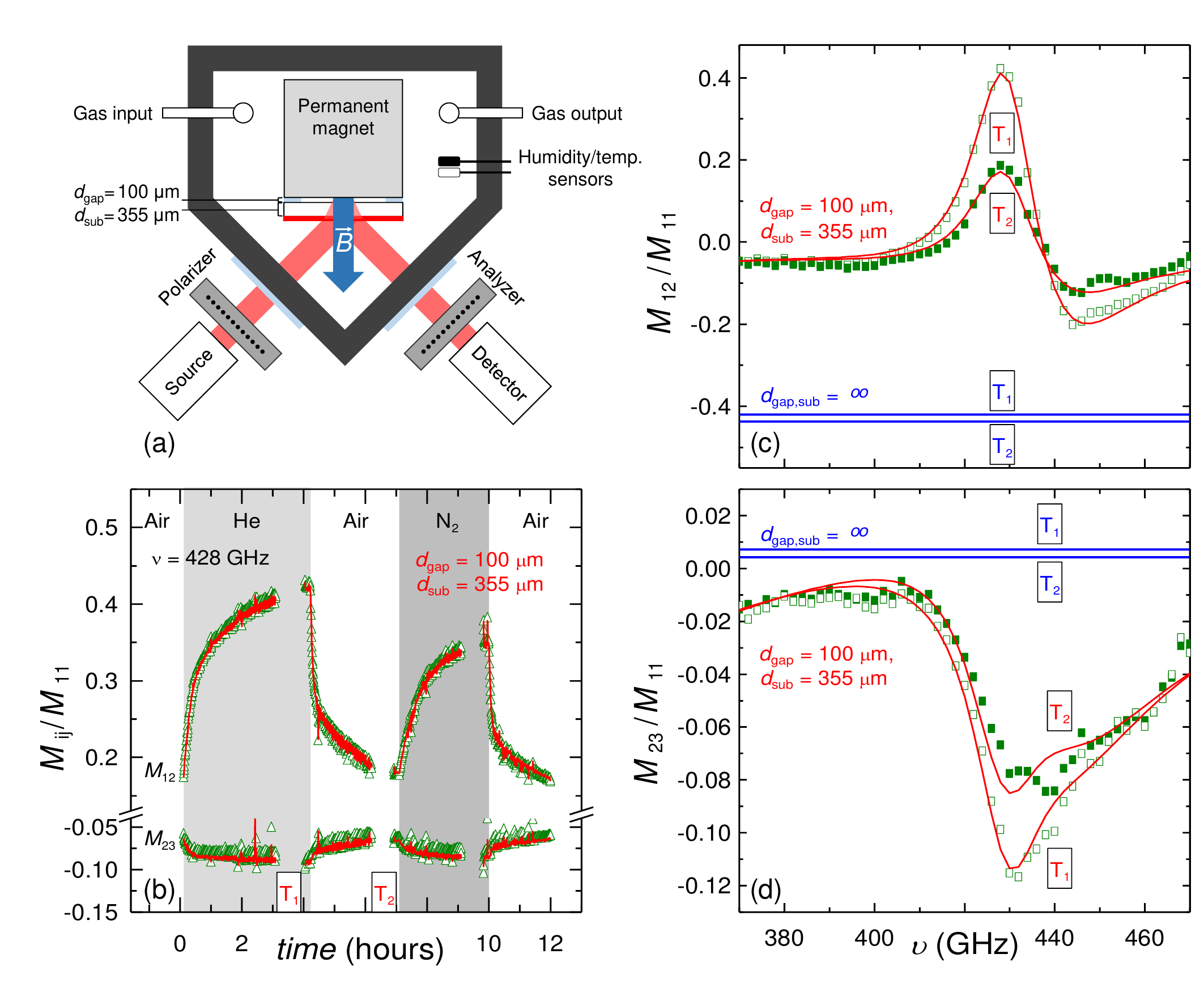}
\caption{\textit{In-situ} cavity-enhanced THz-OHE gas flow experiment and results. Contents of Figs.~\ref{fig:gas-cell-experiments}(a,b) are adapted from Ref.~\onlinecite{knight2017situ} which is licensed under a Creative Commons Attribution 4.0 International License\cite{creative_commons_license_url}. Gas cell schematic (a) for gas exposure experiment on epitaxial graphene grown on SiC. In panel (b), \textit{in-situ} experimental (open triangles) and best-match model data (solid red line) at single frequency ($\nu$ = 428 GHz) for two selected Mueller matrix elements are shown. Normalized $M_{\text{12}}$ (c) and $M_{\text{23}}$ (d) spectra are shown at two different times during the gas exposure experiment (labeled `T$_{1}$' and `T$_{2}$') for the cases with the cavity-enhancement effect ($d_{\text{gap}} = 100~\mu$m and $d_{\text{sub}} = 355~\mu$m) and without ($d_{\text{gap,sub}} = \infty$). In panels (c) and (d) experimental data is shown as open and closed squares, and model-calculated data are solid lines. Note, the experiment is performed only on one side of the permanent magnet (north pole-face) and thus Mueller matrix difference data is not obtained. 
\label{fig:gas-cell-experiments}}
\end{figure*}

Figures~\ref{fig:gas-cell-experiments}(c,d) depict experimental and best-match model data for single-field-orientation cavity-enhanced OHE data as a function of frequency for the graphene sample. These measurements were performed at two different points during the gas exposure experiment: the first spectral measurement was after three hours of exposure to helium, and the subsequent spectral measurement after two hours of exposure to ambient air (labeled `T$_{1}$' and `T$_{2}$' in Fig.~\ref{fig:gas-cell-experiments}(b), respectively). Indicated in Figs.~\ref{fig:gas-cell-experiments}(c,d) are also the THz-OHE data without cavity enhancement (blue lines: $d_{\text{gap,sub}} = \infty$). Comparing the cavity-enhanced data to the case of no cavity-enhancement indicates large changes in the Mueller matrix data are entirely due to interference enhancement in the substrate and external cavity. The off-block-diagonal element $M_{23}$ is selected to show the OHE signature enhancement.

\subsubsection{\textit{In-situ} time-dependent single-frequency single-cavity measurements}

Figure~\ref{fig:gas-cell-experiments}(b) shows the \textit{in-situ} cavity-enhanced THz-OHE data taken at a single frequency, $\nu$ = 428 GHz, for $M_{12}$ and $M_{23}$. Analyzing the data allows the extraction of the graphene's free charge carrier properties $N_\text{s}$, $\mu$, and charge carrier type as a function of time. The carrier type is determined to be \textit{n}-type during each gas phase. It is found that $N_\text{s}$ increases with helium and nitrogen exposure, and decreases with air exposure. An inverse relationship is observed for $\mu$ and $N_\text{s}$ throughout the gas flow experiment. The lowest $N_\text{s}$ occurred at the end of the second air phase where $N_\text{s} = (8.40\pm0.72) \times 10^{11}~\text{cm}^{-2}$ and $\mu = (2595\pm217)~\text{cm}^{2}/\text{Vs}$. The highest $N_\text{s}$ occurred at the end of the helium phase where $N_\text{s} = (2.31\pm0.29) \times 10^{12}~\text{cm}^{-2}$ and $\mu = (1961\pm250)~\text{cm}^{2}/\text{Vs}$. Further details on the \textit{in-situ} THz-OHE gas exposure experiment can be found in our previous publication.\cite{knight2017situ} Sensitivity to the free charge carrier properties as a function of gas flow is entirely dependent on the cavity-enhancement effect, as demonstrated in Figs.~\ref{fig:gas-cell-experiments}(c,d). The variations in free charge carrier properties upon exposure with He and Air, indicated by the blue lines in Figs.~\ref{fig:gas-cell-experiments}(c,d), would not have been detectable with the same instrument without the external cavity stage since the changes in Mueller matrix elements are below the detection limit.   

\section{Conclusion}

We demonstrated a tunable cavity-enhanced THz frequency-domain OHE technique to extract the free charge carrier properties of 2DEG layers situated on top of THz-transparent substrates. A HEMT structure grown on sapphire and epitaxial graphene grown on SiC are studied as examples. For the HEMT structure sample, the OHE signatures are enhanced by tuning an externally coupled Fabry-P\'{e}rot cavity via stepper motor. Data measured as a function of external cavity size and frequency are analyzed to obtain the carrier concentration, mobility, and effective mass parameters of the 2DEG located within the HEMT structure. For the epitaxial graphene on SiC sample, an external cavity of fixed size is used to enhance the OHE signal during a gas flow experiment. This enhancement effect allows the extraction of the graphene's carrier concentration and mobility as a function of time throughout the experiment. In our experiments, the Fabry-P\'{e}rot cavity-enhancement is made possible by the THz-transparent substrates as well as the external cavity (air gap) between the sample's backside and the reflective metal surface. The magnetic field necessary for the OHE experiments is provided by a permanent magnet; for which the metallic coating also provides the reflective surface for the external cavity. Our enhancement technique can be expanded upon by using superconducting magnets to measure samples with much lower free charge carrier contributions. In general, our technique is a powerful method for materials characterization, and can be used to study even more complex sample structures.

\section{Acknowledgments}
We thank Dr.~Craig~M.~Herzinger for helpful discussions. Prof. Nikolas Grandjean is gratefully acknowledged for providing the AlInN/GaN HEMT structure. We thank Dr. Chamseddine Bouhafs and Dr. Vallery Stanishev for growing the graphene sample, and Prof. Rositsa Yakimova for providing access to her sublimation facility for graphene growth. This work was supported in part by the National Science Foundation under award DMR 1808715, by Air Force Office of Scientific Research under award FA9550-18-1-0360, by the Knut and Alice Wallenbergs Foundation supported grant 'Wide-bandgap semiconductors for next generation quantum components', by the Swedish Agency for Innovation Systems under the Competence Center Program Grant No. 2016-05190 and 2014-04712, Swedish Foundation for Strategic Research (RFI14-055, EM16-0024), Swedish Research Council (2016-00889), Swedish Government Strategic Research Area in Materials Science on Functional Materials at Link\"oping University (Faculty Grant SFO Mat LiU No. 2009 00971), and the J.A. Woollam Foundation.


\bibliography{SeansRSIbibFile.bib}

\end{document}